\documentclass[onecolumn]{article}

\usepackage{graphicx}
\usepackage{amsmath}
\usepackage{amssymb}
\usepackage{bigstrut}
\usepackage{tabularx,ragged2e,booktabs,caption,authblk}
\newcolumntype{C}[1]{>{\Centering}m{#1}}

\everymath{\displaystyle}
\usepackage{fullpage}
\usepackage{cite}
\usepackage{float}
\usepackage{url}
\usepackage{hyperref}

\usepackage{tabularx}   
\usepackage{booktabs}   
\usepackage{array}      
\usepackage{float}      

\newcolumntype{L}{>{\raggedright\arraybackslash}X}

\title{An Alignment-Free Explanation for Collective Predator Evasion in Moving Animal Groups}

\author[]{Daniel Str\"{o}mbom\thanks{\mbox{Corresponding author: stroembp@lafayette.edu}}}
\author[]{Catherine Futterman}

\affil[]{Department of Biology, Lafayette College, Easton, PA 18042, USA}

\date{}

\begin{document}

\maketitle

\begin{abstract}
Moving animal groups consist of many distinct individuals but can operate and function as one unit when performing different tasks. Effectively evading unexpected predator attacks is one primary task for many moving groups. The current explanation for predator evasion responses in moving animal groups require the individuals in the groups interact via (velocity) alignment. However, experiments have shown that some animals do not use alignment. This suggests that another explanation for the predator evasion capacity in at least these species is needed. Here we establish that effective collective predator evasion does not require alignment, it can be induced via attraction and repulsion alone. We also show that speed differences between individuals that have directly observed the predator and those that have not influence evasion success and the speed of the collective evasion process, but are not required to induce the phenomenon. Our work here adds collective predator evasion to a number of phenomena previously thought to require alignment interactions that have recently been shown to emerge from attraction and repulsion alone. Based on our findings we suggest experiments and make predictions that may lead to a deeper understanding of not only collective predator evasion, but also collective motion in general.
\end{abstract}

\section*{Introduction}
Despite being made up of many distinct individuals, moving animal groups often appear to operate as one unit \cite{Camazine2001,Sumpter2010}. They are self-organizing aggregates that can assume a variety of group shapes, change group shape, and collectively perform a number of tasks \cite{Shaw1978,Heppner1974,Ward2016,Teddy2016} such as collective decision making \cite{Ward2011}, migration \cite{Flack2018,Couzin2018}, foraging \cite{Evans2019}, and predator evasion \cite{Procaccini2011}. Accomplishing these tasks requires that local information can spread effectively in the group. In particular, information about a predator attack at a specific location in the group must propagate through the group and result in a collective evasive response \cite{Sumpter2008}. Empirical studies in animals across taxa, including birds and fish, have investigated the effect of predator attacks on moving groups and characterized the group response to these. In particular, several studies have shown that the response is mediated through the initiation and spread of escape waves through the group in birds \cite{Procaccini2011,Beauchamp2012}, fish \cite{Radakov1973,Marras2012,Teddy2015} and insects \cite{Treherne1981}. Herbert-Read et al. 2015 conducted experiments where schools of pacific blue-eyes (\textit{Pseudomugil signifer}) moved in an annulus-shaped shallow water tank and when the school approached a particular region in the tank a piston shot out to simulate a predator attack. When undisturbed the school assumed a polarized group configuration and swam around the annulus. When the piston shot out, in most trials, a small proportion of fish closest to it turned and swam directly away from simulated predator at an increased speed. These few fast evading fish then influenced other fish to also turn until the whole school had switched direction and continued to move away from the predator. 

The standard explanation for how moving groups form and function is that each individual interacts locally with its nearby neighbors via some combination of three social forces, attraction, repulsion and (velocity) alignment \cite{Ward2016}. Theoretical studies using so called self-propelled particle (spp) models have shown that a combination of these three forces at the individual level is sufficient to produce a range of stable groups \cite{Aoki1982,Huth1992,Couzin2002,Vicsek2012,Reynolds1987,DOrsogna2006,Romenskyy2017,Cambui2017,Sayama2008}, as well as groups that can perform collective decision making \cite{Couzin2005}, migration, foraging, and predator evasion. In particular, various aspects of collective predator evasion have been studied using models that include alignment interactions \cite{Procaccini2011,Papadopoulou2022b,Papadopoulou2022,Hemelrijk2015,Hildenbrandt2010,Inada2002,Teddy2015}. However, empirical studies have established that several species of fish appear to not rely on alignment to organize their schools \cite{Teddy2011,Katz2011}, so an alternative explanation for this phenomenon in, at least, these species is required.

Recently, a number of studies have shown that several properties of moving animal groups previously thought to require the alignment interaction can be generated without it \cite{Romanczuk2009,Strombom2011,Ferrante2013,Barberis2016,Strombom2019,Strombom2021,Strombom2022front,Strombom2022nerc,Strombom2022dyn}. For example, attraction alone and combinations of attraction and repulsion have been shown sufficient to generate all three standard group types that alignment-based models can generate. In particular, polarized (or aligned) groups \cite{Strombom2011,Strombom2015,Strombom2022front} previously thought to require alignment interactions. Alignment-free models have also been shown capable of generating more disruptive phenomena, e.g. groups that spontaneously transition from one group type to another \cite{Strombom2022nerc,Strombom2022dyn}, as exemplified by schools of golden shiner fish \cite{Tunstrom2013}. Such dynamics had not previously been reported as producible using spp models and had even been conjectured impossible to generate using models of this type due to their inherent averaging of interactions \cite{Katz2011}.

Given the capacity of these recent alignment-free models to generate disruptive phenomena, investigating whether they can provide an attraction and repulsion based explanation for collective predator evasion behavior in moving animal groups should be explored. This investigation is particularly important because several animals are thought to rely on attraction and repulsion alone to organize their groups \cite{Teddy2011,Katz2011,Strombom2022nerc,Strombom2022dyn}. This approach could also yield experimentally testable predictions for predator evasion in animals known to use alignment-free interactions. Additionally, comparing characteristics of alignment-free and alignment-based predator evasion may provide a deeper understanding of this phenomenon across taxa.

\section*{Model and Methods}
\begin{figure}[h!]\centering
\includegraphics[width=10.5 cm]{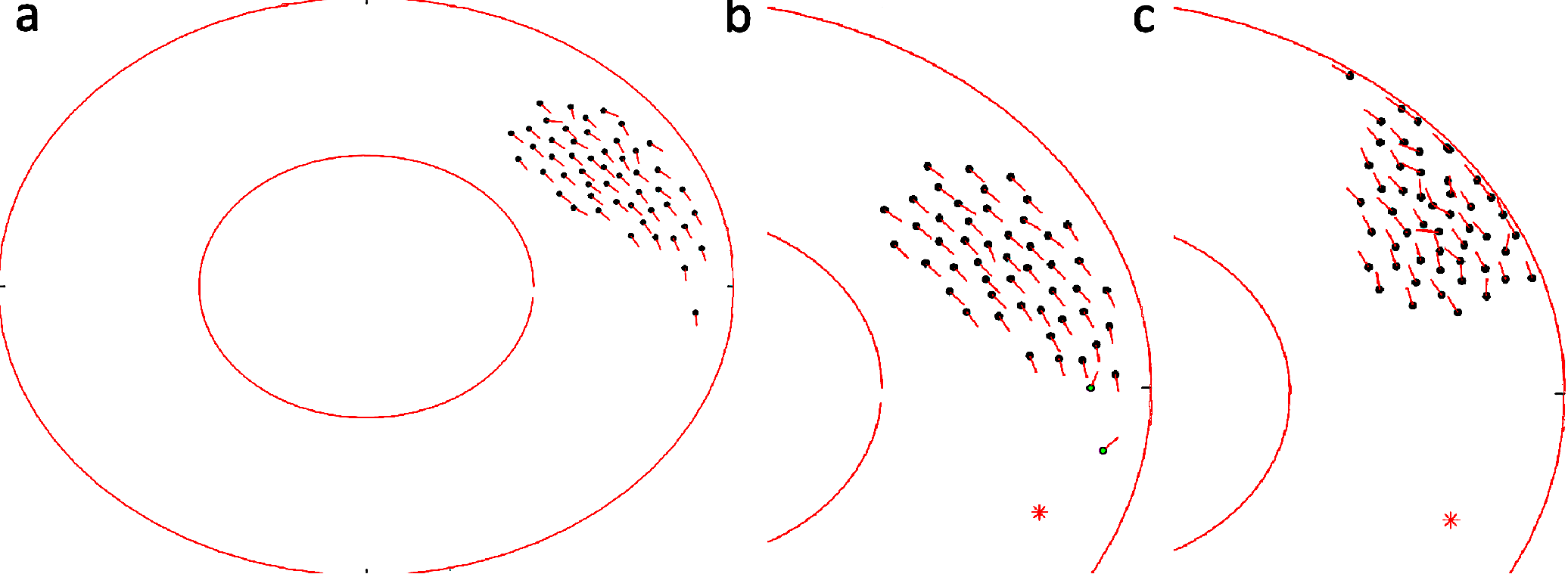}
\caption{Simulation environment. (\textbf{a}) Screenshot of a simulation in the annular region with a group of particles moving clockwise. Black dots represent the particles and the red rods indicate each particles direction of travel. The instantaneous alignment $\phi_t$ of this group at this time $t$ will be positive and large in absolute value (between 0.75 and 1).(\textbf{b}) The simulated predator attack has commenced. The red asterisk indicates the predator attack point and two particles at the front of the group (green dots) have detected the predator and are moving directly away from it. As particles continue to switch from the clockwise to the counter clockwise direction the instantaneous alignment $\phi_t$ will decrease from its positive large value. (\textbf{c}) Following the predator attack the entire group has switched direction and are all moving in the counter clockwise direction away from the predator. The instantaneous alignment $\phi_t$ of this group at this time $t$ will be negative and large in absolute value (between -0.75 and -1). \label{fig1}}
\end{figure}

Here we adapt the self-propelled particle model introduced in \cite{Strombom2015} and subsequently shown capable of generating disruptive phenomena in \cite{Strombom2022nerc,Strombom2022dyn}. In summary, in this model $N$ particles are moving around at constant speed $\delta$. Each particle interacts with other particles within a distance of $R$ from it, excluding those in a blind zone defined by an angle $\beta$ behind the particle relative to its direction of travel. The particles that a particle is interacting with are referred to as its neighbors. On each timestep $t$ each particle determines its neighbors and updates its heading based on attractive and repulsive interactions with its current neighbors. See \cite{Strombom2022nerc,Strombom2022dyn} for a full description of this base model. Here we adapt this model to mimic the setup in the Herbert-Read et al. 2015 experiments. The modifications and changes to the original model are described below.\\
1. The space on which the particles are moving is restricted to an annular region, modelled as two concentric circles with particles confined to move between them. We use the same slip boundary conditions as in \cite{Strombom2022nerc,Strombom2022dyn} modified to work with these circular boundaries.  \\
2. The particles are initialized in a polarized configuration (as observed in experiments) moving clockwise (CW) in the annular region. See Figure \ref{fig1}a.\\
3. As the group of particles approach a specified point with coordinates $(x_p,y_p)$ located inside the annular region ahead of the moving group, a predator attack is simulated at a specified time $t_p$. All particles that are within a distance of $R_p$ from the predator attack point $(x_p,y_p)$ detect the predator when it becomes active at time $t_p$ and are strongly repelled directly away from it. See Figure \ref{fig1}bc. More specifically, if a particle $i$ is within $R_p$ of the predator point its heading update has an additional predator evasion term $\bar{R}^p$ added to Equation 1 in \cite{Strombom2022dyn}. The most general form of the heading update formula used here is \begin{equation}\label{eq:1}\bar{D}_{i,t+1} =  b\hat{D}_{i,t} + c \hat{C}_{i,t} + a \hat{A}_{i,t} + r \bar{F}_{i,t}+ r_p \bar{R}^p_{i,t}+e\hat{\epsilon}_{i,t}, \end{equation} where $\hat{D}_{i,t}$ is particle $i$’s current heading, $\hat{C}_{i,t}$ is the normalized direction towards the local center of mass of its neighbors except those in the blind zone, $\hat{A}_{i,t}$ is the normalized average heading of its neighbors, $\bar{F}_{i,t}$ is a local distance dependent repulsion term, $\bar{R}^p_{i,t}$ is the repulsion directly away from the predator, $\hat{\epsilon}_{i,t}$ is a heading noise term, and the parameters $a$, $b$, $c$, $e$, $r$, $r_p$ specifies the relative strengths of the different interaction terms. For a detailed description of all terms except the predator repulsion term see \cite{Strombom2015,Strombom2022dyn}. The predator repulsion term $\bar{R}^p_{i,t}$ has exactly the same form as the local repulsion term $\bar{F}_{i,t}$, but is calculated using only the predator coordinate $(x_p,y_p)$ instead of the coordinates of all neighbors of the particle.\\
4. To investigate the potential impacts of a difference in speeds between particles that have and those that have not detected the predator we add a parameter $\delta_s$ in some analyses. We denote the ratio of the speed of particles that have detected the predator ($\delta_s$) and the speed of those that have not ($\delta$) by $\Delta=\delta_s/\delta$. So, if there is no difference in speed between those that have detected it and those that have not then $\Delta=1$, if those that have detected it moves twice as fast as those that have not then $\Delta=2$, and so on. 

\subsection*{Measures}
To quantify the simulation results we use two measures similar to those defined and used in \cite{Teddy2015}. The main difference in implementation is that we do not partition the arena into 36 segments for the calculations but instead exclusively use the actual particle positions throughout. \\
1. The arc distance between an individual particle $i$ and the predator point $(x_p,y_p)$ at time $t$, which we denote by $d_{i,t}$. $d_{i,t}$ is calculated as the length of circular arc between the angle of the predator point and the angle of particle $i$ at time $t$, with radius equal to the euclidean distance from fish $i$'s coordinates at time $t$ $(x_{i,t},y_{i,t})$ to the center of the annular region $(0,0)$.\\
2. The instantaneous alignment of the group at time $t$ ($\phi_t$) can be used to determine whether the group is moving clockwise (CW) or counter clockwise (CCW) in the annular region. It is calculated by first calculating the relative orientation of each particle $i$ at time $t$ via $\chi_{i,t}=sin^{-1}(sin(\alpha-\beta))$, where $\alpha$ is the angle of the particle and $\beta$ is the angle of the annular region radius going through the particle position at time $t$. $\chi_{i,t}=-\pi/2$ if the particle is moving CW and $\chi_{i,t}=\pi/2$ if it is moving CCW. The instantaneous alignment of the group at time $t$ is then calculated as the average of the relative orientations of all the $N$ particles at time $t$ through 

\begin{equation}\phi_t=\frac{2}{\pi N}\sum_{i=1}^N \chi_{i,t}.\end{equation}

The instantaneous alignment $\phi_t$ ranges from -1 to 1 and provides information about whether the group is moving clockwise (CW) or counter-clockwise (CCW). An ideal group moving clockwise at time $t$ will have $\phi_t=1$, and one moving counter clockwise will have $\phi_t=-1$. In practice though, because no group, of either real animals or simulated particles, will be moving in an ideal fashion, we classify a group at time $t$ as moving clockwise if $\phi_t\in[0.75,1]$ and counter-clockwise if $\phi_t\in[-0.75,-1]$.

\subsection*{Simulations}
Given that the analyses in \cite{Teddy2015} focused on 51 fish we perform 1450 simulations for each of eight different cases for 51 particles here: four with attraction and repulsion only ($a=0$ in Equation \ref{eq:1}), and four with alignment added ($a=0.5$ in Equation \ref{eq:1}). The four different cases for each correspond to different speed ratios $\Delta =$ 1, 2, 3 and 4. At the start of each simulation particle positions and headings are initialized to be in a polarized configuration moving clockwise. The base configuration for this was obtained by simulating the model from random initial conditions in the annular region without predator attacks until it adopted a polarized configuration moving clockwise. At the start of each simulation this base configuration is then perturbed both with respect to particle positions and headings to obtain unique initial conditions for each of the 1450x8=11600 simulations carried out. The predator attack point $(x_p,y_p)$ was chosen to be at coordinates $(17,-7)$ because with this value a few particles at the front of the moving group tend to be within $R_p$ from the predator point $(x_p,y_p)$ at the predator attack time $t_p=35$. These few particles will detect the predator and be assigned the higher speed $\delta_s$ corresponding to that particular case. The other parameters in the heading update (Equation \ref{eq:1}) were set to $c=2$, $d=0.5$, $r=1.5$, $r_p=5$, $\delta=0.2$ and $e=1/14$. The blind angle $\beta=3.2$ was chosen based on the findings in \cite{Strombom2022dyn}. At this value, a group tends to adopt a polarized configuration, but is close enough to the bistability region to reconfigure in response to perturbations. In each of the 11600 simulations we collect the trajectory of each particle over time. This trajectory information was then used to present representative simulations of the key cases we seek to compare, i.e. collective predator evasive behavior generated by attraction and repulsion alone without any speed difference ($\Delta=1$) and with the largest speed difference used ($\Delta=4$), and the same speed differences when alignment was added. Using the trajectories of all particle through each simulation we calculated the distance between each particle and the predator attack point through time and plotted the result in Figure \ref{fig2}. The predator attack time $t_p=35$ and predator detection radius $R_p=8$ was also superimposed on each plot. This trajectory information was also used to calculate the instantaneous alignment over time in each simulation for each case. The instantaneous alignment was then used to calculate the median over the 1450 simulations in each of the eight cases to generate the average evasion curves in Figure \ref{fig3}, as well as the mean, and 95\% confidence intervals to generate supplementary Figures S1 and S2. Due to non-normal instantaneous alignment measurement distributions all confidence intervals presented are bootstrap confidence intervals calculated using the Matlab function bootci with 2000 bootstrap samples.
To investigate how evasion success depends on the interaction between the interaction rules and speed differences we ran 7000 simulations for each $\Delta$ from 1 to 4 in increments of 0.1 for each of the interaction rules and measured the instantaneous alignment over time. This information was then used to calculate the mean proportion of simulations, as well as 95\% confidence intervals, in each of the 62 cases (2 interaction rules and 31 speed differences) that resulted in a successful group evasive manoeuvre away from the predator. Using the criteria that if $\phi_t< -0.75$ at some time $t$ after the predator attack time $t_p$, then the group turned away from the predator and is moving away from it counter clockwise. The results of these analyses were used to create Figure \ref{fig4} and Table \ref{tab1}.
 See the Data Availability Statement for information about how to access the Matlab code used for these analyses.

\section*{Results}
\begin{figure}[h!]\centering
\includegraphics[width=10.5 cm]{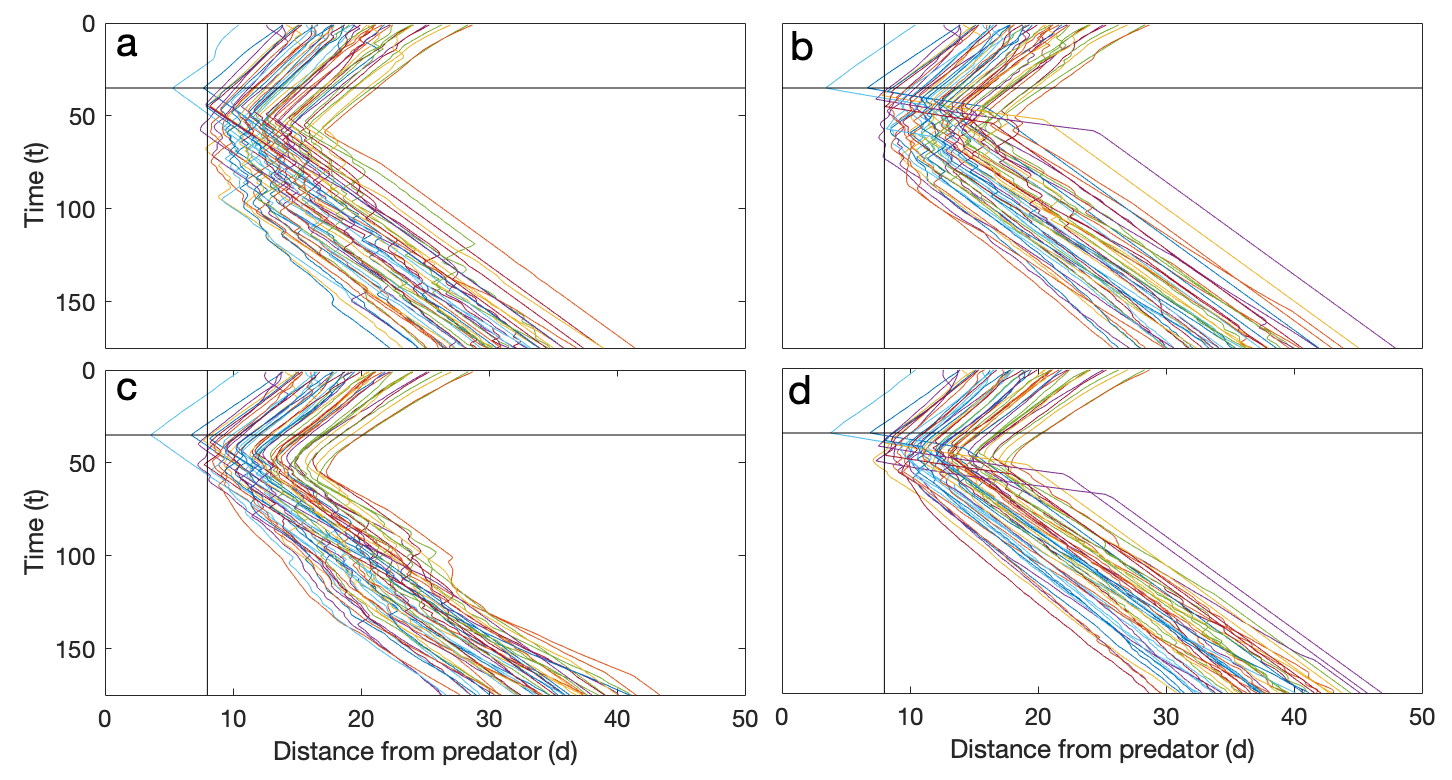}
\caption{The collective evasion process. Each panel shows the angular distance of each particle to the predator point over time. The horizontal line indicates the predator time $t_p=35$ and the vertical line indicated the predator detection range $R_p=8$. In each case we see a similar process. The distance between all fish and the predator decreases up until the predator attack at $t=35$ at which point the particles that are closer to the predator than $R_p=8$ (left of the vertical line) starts moving away from the predator and influences other particles (which have not detected the predator) to switch their directions also in a cascade type fashion. The two top panels (\textbf{a,b}) shows the process when only attraction and repulsion are operating with no speed difference $\Delta=1$ (\textbf{a}) and with a large speed difference $\Delta=4$ (\textbf{b}). The two bottom panels (\textbf{c,d}) show the process when alignment is operating, with no speed difference $\Delta=1$ (\textbf{c}) and with a large speed difference $\Delta=4$ (\textbf{d}). By comparing the panels on the same row, we can observe the effects of differing speeds, and by the panels in the same column we can see the effects of including alignment. While there are differences, in particular, that higher speed differences result in smoother transitions and more stable group motion post evasion maneuver, the basic evasion behavior is the same.  \label{fig2}}
\end{figure}

Alignment interactions are not required to induce collective predator evasion in moving groups. Attraction and repulsion alone are sufficient. Figures \ref{fig2}ab shows the distance $d$ from the predator to each particle over time in the alignment-free model, and Figures \ref{fig2}cd show the corresponding trajectories in the alignment-based model. In all cases we note that all particles approach the predator location, the predator is activated at $t=35$ and at this time the few particles that are within a distance of $R_p=8$ from the predator move directly away from it and influences the other particles, which have not detected the predator, to turn until the whole group has turned and collectively move away from the predator. While there are noticeable differences between the trajectories in the different panels the basic dynamics are the same, implying that alignment is not strictly required to induce this phenomenon.

Speed differences between individuals that detect the predator and those that do not influence the success rate of and stabilize the predator evasion process, but are not strictly required for it to occur. Comparing Figure \ref{fig2}a with \ref{fig2}b, and figures \ref{fig2}c with \ref{fig2}d we note that the post attack trajectories smooth out faster when there is a large speed difference $\Delta$. This implies that the evasion process is faster and the whole group switches to its new direction in a smoother fashion when speed differences are large.

\begin{figure}[h!]\centering
\includegraphics[width=10.5 cm]{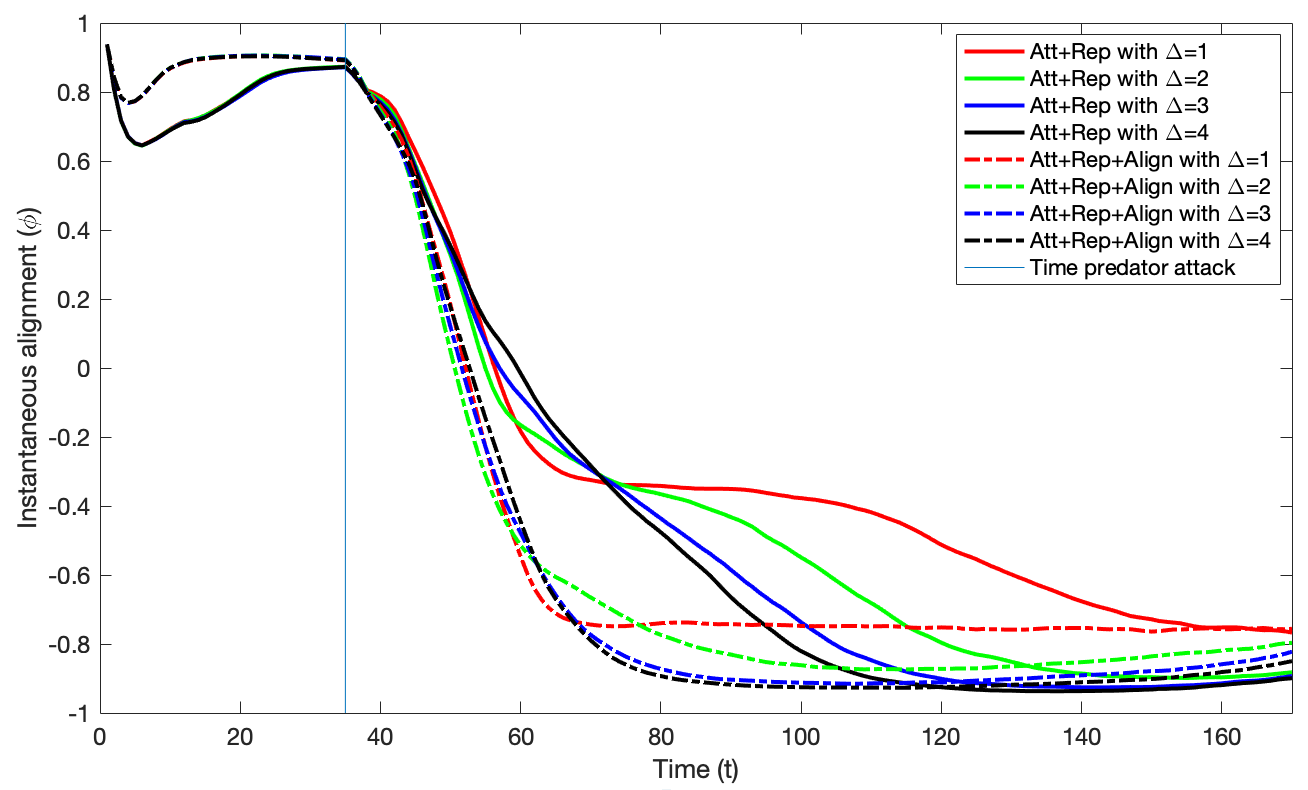}
\caption{Comparison of the average instantaneous alignment ($\phi$) over time for the eight different cases. Solid curves correspond to attraction and repulsion only, dashed lines correspond to alignment included. Curve color indicates the speed difference $\Delta=1$ (red), $\Delta=2$ (green), $\Delta=3$ (blue), $\Delta=4$ (black). The horizontal line represents the predator attack time $t_p=35$ and we note that at this time all curves drop rapidly from a high $\phi\approx0.9$. Initially all curves drop at comparable rates but as time progresses the alignment curves drop quicker, especially when the $\phi$ drops below -0.2, i.e. after the average heading of the group has turned away from the direction of the predator. We also note that these are averages and that the successful turn rate is significantly lower for the smaller speed differences than the higher (Table \ref{tab1}), which contributes to the flattening out of the green and red curves (due to failure to turn, splits etc). See Video S1 for example simulations of each of the eight cases.  \label{fig3}}
\end{figure}

Figure \ref{fig3} shows the average evasion behavior over time for all the eight cases, and Video S1 shows example simulations of each. We note that, on average, there are noticeable differences between evasive behavior generated by alignment-free (solid lines) and alignment-based (dashed lines) models in Figure \ref{fig3}. While they are relatively similar up until $\phi\approx-0.2$, i.e. when there is a net instantaneous alignment in the direction away from the predator but a significant part of the group has yet to turn, after that the behavior of two model types diverge. Alignment-based models exhibit a smooth switch on average whereas the alignment-free models exhibit a slower turn rate after $\phi\approx-0.2$. We also note that speed difference plays a significant role on the average behavior. In particular, the higher speed difference cases ($\Delta=3$ and $4$) for both model types exhibit smooth decreases to low phi values (<-0.75), indicating successful evasions, on average. The low speed difference curves ($\Delta=1$ and $2$) exhibit a different type of interrupted turning behavior, on average. This is a consequence of the fact that the lower speed difference cases have lower evasion success rate and are more susceptible to failing to turn and pass the predator or experience group splits. See supplementary materials for versions of Figure \ref{fig3} with mean and 95\% confidence interval information (Figure S1) and the raw instantaneous alignment measurements over all simulations (Figure S2). The proportion of successful evasions, including 95\% confidence interval information, as a function of speed difference $\Delta$ for both interaction rules is presented in Figure \ref{fig4}. The green curve shows the proportion of success for attraction and repulsion only, and the red curve for attraction, repulsion and alignment. The red curve shows that the alignment-based model has a success rate of above 0.95 for all speeds, and the green curve shows that the alignment-free model's evasion success increases from about 0.65 at no speed difference ($\Delta=1$) to about 0.97 for $\Delta=4$. Overall, in both cases the success proportion increases with increasing speed difference. Table \ref{tab1} shows the specific proportions for integer speed differences. We note that for large $\Delta$ values the success proportions are high for both interaction rules, that alignment-based models have higher success rate overall, and that the alignment-free model with no speed difference only has a 64\% success percentage, which explains the flattening of the red-solid curve after -0.2.

\begin{figure}[h!]\centering
\includegraphics[width=10.5 cm]{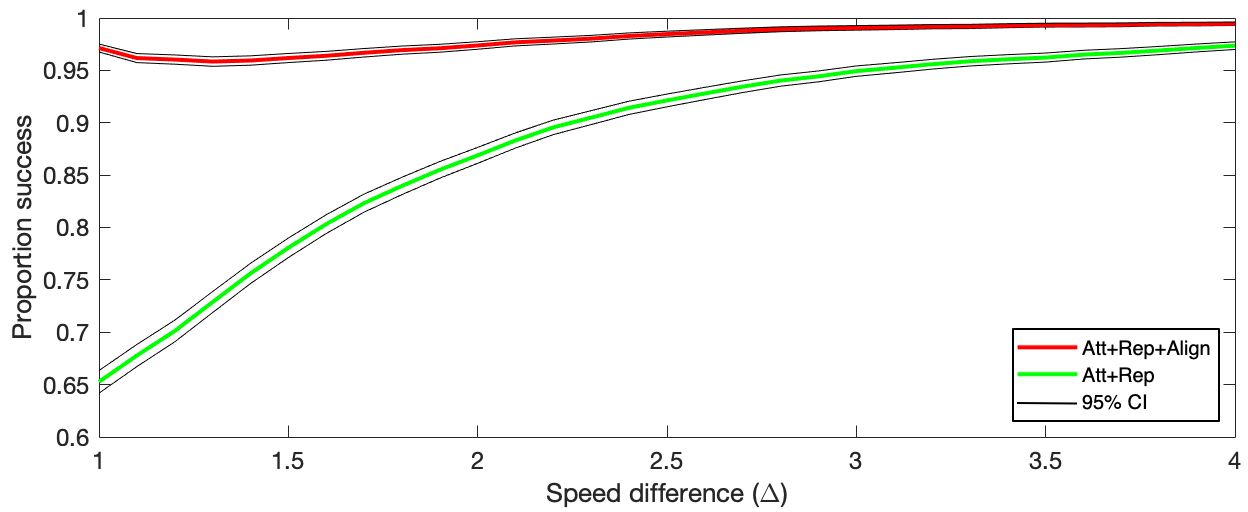}
\caption{Evasion success as a function of speed difference. The green curve shows the mean evasion success proportion for the attraction and repulsion only model, and the red curve the mean evasion success proportion for the attraction, repulsion and alignment model. The black lines represent the 95\% bootstrap confidence intervals. \label{fig4}}
\end{figure} 

\begin{table}[ht]
\centering
\caption{Probability of successful evasion depending on the interaction rule used and speed differences between particles that have detected the predator and those that have not. We note that increasing speed differences ($\Delta$) increase evasion success for both interaction rules.}
\label{tab1}
\begin{tabular}{lcccc}
\hline
& $\Delta=1$ & $\Delta=2$ & $\Delta=3$ & $\Delta=4$ \\
\hline
Att+Rep         & 0.6527 & 0.8728 & 0.9483 & 0.9736 \\
Att+Rep+Align   & 0.9713 & 0.9744 & 0.9907 & 0.9945 \\
\hline
\end{tabular}
\end{table}

\section*{Discussion}
Alignment interactions are not required to induce collective predator evasion in moving groups. This finding provides an explanation for collective predator evasion in moving animal groups for animals that do not use alignment interactions to organize their groups. This extends our mechanistic understanding of this phenomena beyond the alignment-based explanation provided by the models in \cite{Papadopoulou2022b,Papadopoulou2022,Hemelrijk2015,Hildenbrandt2010,Inada2002,Teddy2015}.

Speed differences between individuals that detect the predator and those that do not influence the success rate of, and stabilize the evasion process, but are not strictly required for it to occur (Figures \ref{fig2}, \ref{fig3} and \ref{fig4}). In \cite{Teddy2015} alignment and speed differences were isolated as primarily responsible for the model's predator evasion capacity via escape waves consistent with data. While our findings indicate that predator evasion and escape wave-like behavior does not strictly depend on either of these, we do find that both alignment and speed differences improve the rate and smoothness of the response (Figure \ref{fig2} and \ref{fig3}). This suggests that despite not being technically required they may be highly relevant to predator evasion in some real moving animal groups. In particular, because some fish have been shown to use alignment, e.g. barred flagtails \textit{Kuhlia mugil} \cite{Gautrais2012}, and speed changes during attacks are well documented across taxa \cite{Radakov1973,Michaelsen2002,OBrien1987,Procaccini2011,Beauchamp2012,Treherne1981}.

Our work makes a number of predictions that could be experimentally tested by re-running the experiments performed in \cite{Teddy2015} with fish that are known to use alignment, and fish that are known to not use alignment. Suitable candidates for lab experiments with fish that do not use alignment are golden shiners \textit{Notemigonus crysoleucas} \cite{Katz2011} and mosquitofish \textit{Gambusia holbrooki} \cite{Teddy2011}, and a fish that uses alignment is barred flagtails \textit{Kuhlia mugil} \cite{Gautrais2012}. The results presented in Figure \ref{fig3} predict that fish that rely on alignment and those that do not will exhibit similar responses early on in the process after the attack, but that fish that do not use alignment will be slower in completing the latter part of the process as compared to fish that rely on alignment. For fish using alignment we predict that there will be no noticeable slowing of the process midway. Furthermore, measuring the speed difference between fish that detect the predator and those that do not also yields predictions of the relative success rate of evasions once it is known whether alignment is used or not (Figure \ref{fig4} and Table \ref{tab1}). 

In addition, perhaps the observed escape wave speed differences in three key studies of the phenomena in fish \cite{Radakov1973,Marras2012,Teddy2015} may be partially explained by the different species using different interaction rules. In particular, whether they use alignment or not, and what the speed difference between the primary predator detectors and non-detectors are. The average escape wave speeds found in the three studies where 0.29 m/s \cite{Teddy2015}, 4.1–10.3 m/s \cite{Marras2012}, 11.8–15.1 m/s \cite{Radakov1973}. We are unable to find any information about which interaction rules the three species used in these studies are thought to use, in particular, if either of them have been shown to rely on alignment. While there may be other explanations for this discrepancy in speed, such as size differences in the fish used as pointed out in \cite{Teddy2015}, the rules employed by the different species may also influence this. If hypothetically no other factor is responsible, based on our results we would guess that the pacific blue-eyes used in \cite{Teddy2015} rely on attraction and repulsion only, and the fish in \cite{Marras2012} and \cite{Radakov1973} both rely on alignment. Testing whether pacific blue-eyes rely on alignment or not should be straightforward, given that the trajectory data exists, and the methods used to determine whether alignment is operating in golden shiners, mosquitofish and barred flagtails should be available from \cite{Teddy2011,Katz2011,Gautrais2012}.

Performing these kinds of experiments for a range of species might also be useful in the other direction, i.e. to infer interaction rules. It is known that inferring interaction rules from data of groups in a steady state is less informative than inferences based on data from the approach to the stable state \cite{Mann2011}. Immediately following the predator attacks in \cite{Radakov1973,Teddy2015} the group is out of steady state and recovering towards one over time. This phase could provide particularly useful data for inferring, or at least distinguishing between proposed, interaction rules by the approach described in paragraph 3 of the discussion in \cite{Strombom2022front}.

Here we have shown that collective predator evasion is yet another phenomenon previously thought to require alignment interactions that do not. In particular, the discovery that polarized (or aligned) groups can be generated from attraction alone in combination with a range of biologically plausible auxiliary locomotion related assumptions \cite{Strombom2022front}, including asynchrony \cite{Strombom2011,Strombom2019}, anticipation \cite{Strombom2021}, and burst-and-glide dynamics \cite{Strombom2022front}. Combinations of attraction and repulsion alone have also been shown to have the capacity to generate polarized (or aligned) groups in a range of fundamentally different types of models \cite{Romanczuk2009,DOrsogna2006,Ferrante2013,Barberis2016}. Noting further that the above mentioned models not only can produce polarized (or aligned) groups without including an alignment interaction, but almost all of them have also been shown capable of generating the other two standard groups of collective motion, i.e. mills/tori and disorganized swarms. This indicates that alignment is not required to explain general features of stable collective motion in moving animal groups. However, until recently it was unclear (and actually conjectured impossible on theoretical grounds \cite{Katz2011}), whether any spp-model, alignment-based or otherwise, could generate disruptive phenomena such as bistability and transitioning between groups types. Interestingly, it was recently established that both alignment-free and alignment-based models can actually generate these types of disruptive behaviors \cite{Strombom2022nerc,Strombom2022dyn}. Through the current work we can now add collective predator evasion to the list of phenomena that are explainable by attraction and repulsion alone without alignment required. Through these findings over the past decade it is becoming increasingly plausible that the observed group level alignment, and other phenomena, in many real moving animal groups emerges from an interplay of attraction and repulsion alone, rather than from an explicit alignment interaction. This challenges the current dogma of collective motion which states that "Attraction is a prerequisite for the formation of social aggregations, repulsion restricts crowding and prevents collisions and alignment is proposed to produce coordinated motion." \cite{Ward2016}.

\section*{Data Availability Statement}
The data and Matlab code needed to regenerate all figures and tables in this manuscript and run new simulations can be found at https://github.com/danielstrombom/Evasion.


\bibliographystyle{unsrt}

\begin{thebibliography}{10}

\bibitem{Camazine2001}
Scott Camazine, Jean-Louis Deneubourg, Nigel~R. Franks, James Sneyd, Guy Theraula, and Eric Bonabeau.
\newblock {\em Self-Organization in Biological Systems}.
\newblock Princeton University Press, Princeton, 2001.

\bibitem{Sumpter2010}
D.~J.~T. Sumpter.
\newblock {\em Collective animal behavior}.
\newblock Princeton University Press, 2010.

\bibitem{Shaw1978}
Evelyn Shaw.
\newblock Schooling fishes: the school, a truly egalitarian form of organization in which all members of the group are alike in influence, offers substantial benefits to its participants.
\newblock {\em American Scientist}, 66(2):166--175, 1978.

\bibitem{Heppner1974}
Frank~H Heppner.
\newblock Avian flight formations.
\newblock {\em Bird-Banding}, 45(2):160--169, 1974.

\bibitem{Ward2016}
Ashley Ward and Mike Webster.
\newblock {\em Sociality: the behaviour of group-living animals}.
\newblock Springer, 2016.

\bibitem{Teddy2016}
James~E Herbert-Read.
\newblock Understanding how animal groups achieve coordinated movement.
\newblock {\em Journal of Experimental Biology}, 219(19):2971--2983, 2016.

\bibitem{Ward2011}
Ashley~JW Ward, James~E Herbert-Read, David~JT Sumpter, and Jens Krause.
\newblock Fast and accurate decisions through collective vigilance in fish shoals.
\newblock {\em Proceedings of the National Academy of Sciences}, 108(6):2312--2315, 2011.

\bibitem{Flack2018}
Andrea Flack, M{\'a}t{\'e} Nagy, Wolfgang Fiedler, Iain~D Couzin, and Martin Wikelski.
\newblock From local collective behavior to global migratory patterns in white storks.
\newblock {\em Science}, 360(6391):911--914, 2018.

\bibitem{Couzin2018}
Iain~D Couzin.
\newblock Collective animal migration.
\newblock {\em Current Biology}, 28(17):R976--R980, 2018.

\bibitem{Evans2019}
Julian~C Evans, Colin~J Torney, Stephen~C Votier, and Sasha~RX Dall.
\newblock Social information use and collective foraging in a pursuit diving seabird.
\newblock {\em PloS one}, 14(9):e0222600, 2019.

\bibitem{Procaccini2011}
Andrea Procaccini, Alberto Orlandi, Andrea Cavagna, Irene Giardina, Francesca Zoratto, Daniela Santucci, Flavia Chiarotti, Charlotte~K Hemelrijk, Enrico Alleva, Giorgio Parisi, et~al.
\newblock Propagating waves in starling, sturnus vulgaris, flocks under predation.
\newblock {\em Animal behaviour}, 82(4):759--765, 2011.

\bibitem{Sumpter2008}
David Sumpter, Jerome Buhl, Dora Biro, and Iain Couzin.
\newblock Information transfer in moving animal groups.
\newblock {\em Theory in biosciences}, 127:177--186, 2008.

\bibitem{Beauchamp2012}
Guy Beauchamp.
\newblock Flock size and density influence speed of escape waves in semipalmated sandpipers.
\newblock {\em Animal Behaviour}, 83(4):1125--1129, 2012.

\bibitem{Radakov1973}
Dmitri{\u\i}~Viktorovich Radakov.
\newblock {\em Schooling in the ecology of fish}.
\newblock John Wiley \& Sons, 1973.

\bibitem{Marras2012}
Stefano Marras, Robert~S Batty, and Paolo Domenici.
\newblock Information transfer and antipredator maneuvers in schooling herring.
\newblock {\em Adaptive Behavior}, 20(1):44--56, 2012.

\bibitem{Teddy2015}
James~E Herbert-Read, Jerome Buhl, Feng Hu, Ashley~JW Ward, and David~JT Sumpter.
\newblock Initiation and spread of escape waves within animal groups.
\newblock {\em Royal Society open science}, 2(4):140355, 2015.

\bibitem{Treherne1981}
John~E Treherne and William~A Foster.
\newblock Group transmission of predator avoidance behaviour in a marine insect: the trafalgar effect.
\newblock {\em Animal Behaviour}, 29(3):911--917, 1981.

\bibitem{Aoki1982}
I.~Aoki.
\newblock A simulation study on the schooling mechanism in fish.
\newblock {\em Bull. Jpn. Soc. Fish.}, 48, 1982.

\bibitem{Huth1992}
Andreas Huth and Christian Wissel.
\newblock The simulation of the movement of fish schools.
\newblock {\em Journal of theoretical biology}, 156(3):365--385, 1992.

\bibitem{Couzin2002}
I.~D. Couzin, J.~Krause, R.~James, G.~D. Ruxton, and N.~R. Franks.
\newblock Collective memory and spatial sorting in animal groups.
\newblock {\em J. Theor. Biol.}, 218:1--11, 2002.

\bibitem{Vicsek2012}
T.~Vicsek and A.~Zafeiris.
\newblock Collective motion.
\newblock {\em Physics Reports}, 517:71--140, 2012.

\bibitem{Reynolds1987}
C.~W. Reynolds.
\newblock Flocks, herds and schools: A distributed behavioral model.
\newblock {\em SIGGRAPH Comput Graph}, 21:25--34, 1987.

\bibitem{DOrsogna2006}
Maria~R D'Orsogna, Yao-Li Chuang, Andrea~L Bertozzi, and Lincoln~S Chayes.
\newblock Self-propelled particles with soft-core interactions: patterns, stability, and collapse.
\newblock {\em Physical review letters}, 96(10):104302, 2006.

\bibitem{Romenskyy2017}
Maksym Romenskyy, James~E Herbert-Read, Ashley~JW Ward, and David~JT Sumpter.
\newblock Body size affects the strength of social interactions and spatial organization of a schooling fish (pseudomugil signifer).
\newblock {\em Open Science}, 4(4):161056, 2017.

\bibitem{Cambui2017}
Dor{\'\i}lson~S Cambui.
\newblock Collective behavior states in animal groups.
\newblock {\em Modern Physics Letters B}, 31(06):1750054, 2017.

\bibitem{Sayama2008}
Jonathan~P Newman and Hiroki Sayama.
\newblock Effect of sensory blind zones on milling behavior in a dynamic self-propelled particle model.
\newblock {\em Physical Review E}, 78(1):011913, 2008.

\bibitem{Couzin2005}
Iain~D Couzin, Jens Krause, Nigel~R Franks, and Simon~A Levin.
\newblock Effective leadership and decision-making in animal groups on the move.
\newblock {\em Nature}, 433(7025):513--516, 2005.

\bibitem{Papadopoulou2022b}
Marina Papadopoulou, Hanno Hildenbrandt, Daniel~WE Sankey, Steven~J Portugal, and Charlotte~K Hemelrijk.
\newblock Emergence of splits and collective turns in pigeon flocks under predation.
\newblock {\em Royal Society Open Science}, 9(2):211898, 2022.

\bibitem{Papadopoulou2022}
Marina Papadopoulou, Hanno Hildenbrandt, Daniel~WE Sankey, Steven~J Portugal, and Charlotte~K Hemelrijk.
\newblock Self-organization of collective escape in pigeon flocks.
\newblock {\em PLoS computational biology}, 18(1):e1009772, 2022.

\bibitem{Hemelrijk2015}
Charlotte~K Hemelrijk, Lars van Zuidam, and Hanno Hildenbrandt.
\newblock What underlies waves of agitation in starling flocks.
\newblock {\em Behavioral ecology and sociobiology}, 69:755--764, 2015.

\bibitem{Hildenbrandt2010}
Hanno Hildenbrandt, Cladio Carere, and Charlotte~K Hemelrijk.
\newblock Self-organized aerial displays of thousands of starlings: a model.
\newblock {\em Behavioral Ecology}, 21(6):1349--1359, 2010.

\bibitem{Inada2002}
Yoshinobu Inada and Keiji Kawachi.
\newblock Order and flexibility in the motion of fish schools.
\newblock {\em Journal of theoretical Biology}, 214(3):371--387, 2002.

\bibitem{Teddy2011}
James~E Herbert-Read, Andrea Perna, Richard~P Mann, Timothy~M Schaerf, David~JT Sumpter, and Ashley~JW Ward.
\newblock Inferring the rules of interaction of shoaling fish.
\newblock {\em Proceedings of the National Academy of Sciences}, 108(46):18726--18731, 2011.

\bibitem{Katz2011}
Yael Katz, Kolbj{\o}rn Tunstr{\o}m, Christos~C Ioannou, Cristi{\'a}n Huepe, and Iain~D Couzin.
\newblock Inferring the structure and dynamics of interactions in schooling fish.
\newblock {\em Proceedings of the National Academy of Sciences}, 108(46):18720--18725, 2011.

\bibitem{Romanczuk2009}
Pawel Romanczuk, Iain~D Couzin, and Lutz Schimansky-Geier.
\newblock Collective motion due to individual escape and pursuit response.
\newblock {\em Physical Review Letters}, 102(1):010602, 2009.

\bibitem{Strombom2011}
D.~Str{\"{o}}mbom.
\newblock Collective motion from local attraction.
\newblock {\em J. Theor. Biol.}, 283:145--151, 2011.

\bibitem{Ferrante2013}
Eliseo Ferrante, Ali~Emre Turgut, Marco Dorigo, and Cristi{\'a}n Huepe.
\newblock Elasticity-based mechanism for the collective motion of self-propelled particles with springlike interactions: A model system for natural and artificial swarms.
\newblock {\em Physical review letters}, 111(26):268302, 2013.

\bibitem{Barberis2016}
Lucas Barberis and Fernando Peruani.
\newblock Large-scale patterns in a minimal cognitive flocking model: incidental leaders, nematic patterns, and aggregates.
\newblock {\em Physical review letters}, 117(24):248001, 2016.

\bibitem{Strombom2019}
Daniel Str{\"o}mbom, Tasnia Hassan, W~Hunter~Greis, and Alice Antia.
\newblock Asynchrony induces polarization in attraction-based models of collective motion.
\newblock {\em Royal Society open science}, 6(4):190381, 2019.

\bibitem{Strombom2021}
Daniel Str{\"o}mbom and Alice Antia.
\newblock Anticipation induces polarized collective motion in attraction based models.
\newblock {\em Northeast Journal of Complex Systems (NEJCS)}, 3(1):Article 2, 2021.

\bibitem{Strombom2022front}
Daniel Str{\"o}mbom and Grace Tulevech.
\newblock Attraction vs. alignment as drivers of collective motion.
\newblock {\em Frontiers in Applied Mathematics and Statistics}, 7:717523, 2022.

\bibitem{Strombom2022nerc}
Daniel Str{\"o}mbom, Stephanie Nickerson, Catherine Futterman, Alyssa DiFazio, Cameron Costello, and Kolbj{\o}rn Tunstr{\o}m.
\newblock Bistability and switching behavior in moving animal groups.
\newblock {\em Northeast Journal of Complex Systems (NEJCS)}, 4(1):1, 2022.

\bibitem{Strombom2022dyn}
Daniel Str{\"o}mbom, Grace Tulevech, Rachel Giunta, and Zachary Cullen.
\newblock Asymmetric interactions induce bistability and switching behavior in models of collective motion.
\newblock {\em Dynamics}, 2(4):462--472, 2022.

\bibitem{Strombom2015}
Daniel Str{\"o}mbom, Mattias Siljestam, Jinha Park, and David~JT Sumpter.
\newblock The shape and dynamics of local attraction.
\newblock {\em The European Physical Journal Special Topics}, 224(17-18):3311--3323, 2015.

\bibitem{Tunstrom2013}
Kolbj{\o}rn Tunstr{\o}m, Yael Katz, Christos~C Ioannou, Cristi{\'a}n Huepe, Matthew~J Lutz, and Iain~D Couzin.
\newblock Collective states, multistability and transitional behavior in schooling fish.
\newblock {\em PLoS Comput Biol}, 9(2):e1002915, 2013.

\bibitem{Gautrais2012}
Jacques Gautrais, Francesco Ginelli, Richard Fournier, St{\'e}phane Blanco, Marc Soria, Hugues Chat{\'e}, and Guy Theraulaz.
\newblock Deciphering interactions in moving animal groups.
\newblock {\em Plos computational biology}, 8(9):e1002678, 2012.

\bibitem{Michaelsen2002}
Tore~Christian Michaelsen and Ingvar Byrkjedal.
\newblock 'magic carpet'flight in shorebirds attacked by raptors on a migrational stopover site.
\newblock {\em Ardea}, 90(1):167--171, 2002.

\bibitem{OBrien1987}
DP~O’Brien.
\newblock Description of escape responses of krill (crustacea: Euphausiacea), with particular reference to swarming behavior and the size and proximity of the predator.
\newblock {\em Journal of Crustacean Biology}, 7(3):449--457, 1987.

\bibitem{Mann2011}
Richard~P Mann.
\newblock Bayesian inference for identifying interaction rules in moving animal groups.
\newblock {\em PloS one}, 6(8):e22827, 2011.

\end{thebibliography}

\section*{Supplementary Materials}

\begin{figure}[h!]\centering
\includegraphics[width=10.5 cm]{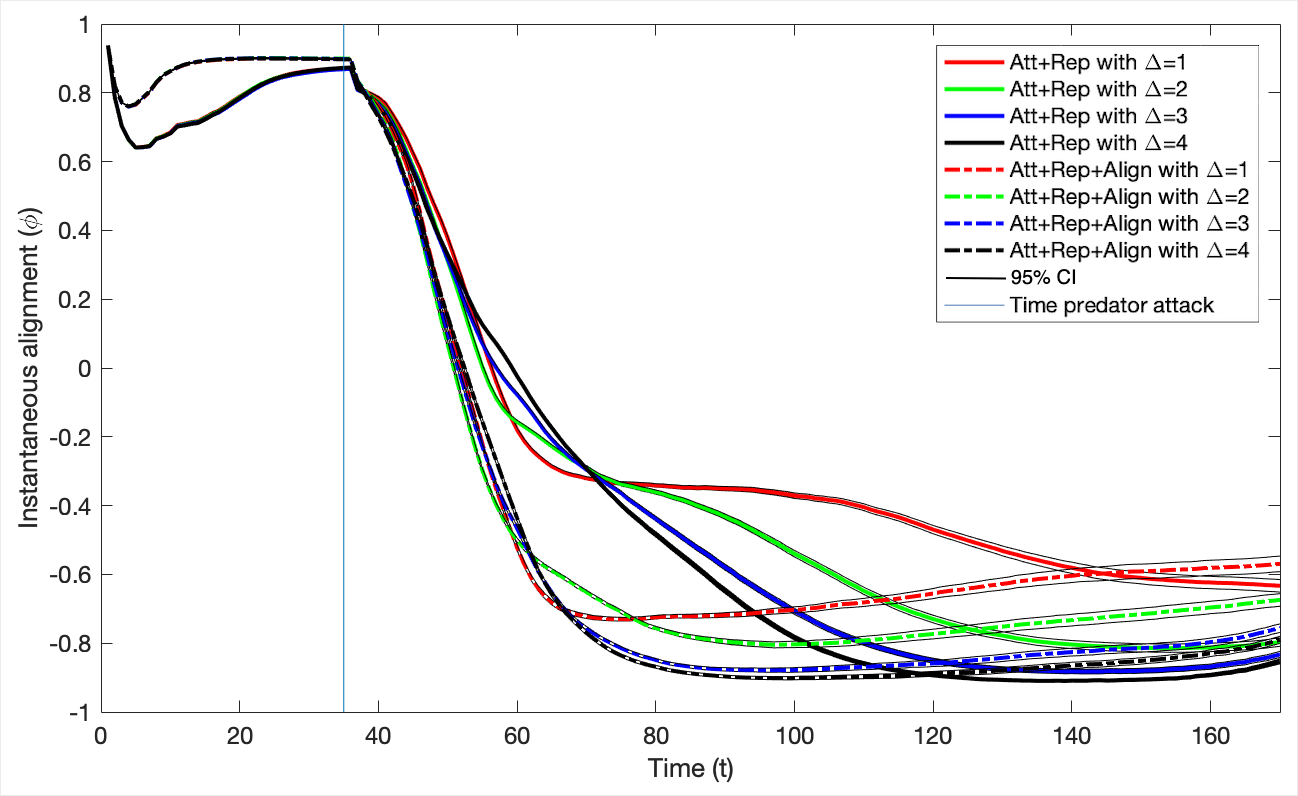}
\caption{Figure S1: Mean and 95\% bootstrap confidence intervals of the instantaneous alignment $\phi$ over the 1450 simulation for each of the eight cases. The black thin curves around each colored curve represents the 95\% confidence intervals. We note that in all cases they are very narrow, sometimes so narrow that they are difficult to distinguish from the colored curves. \label{fig4}}
\end{figure} 

\begin{figure}[h!]\centering
\includegraphics[width=15 cm]{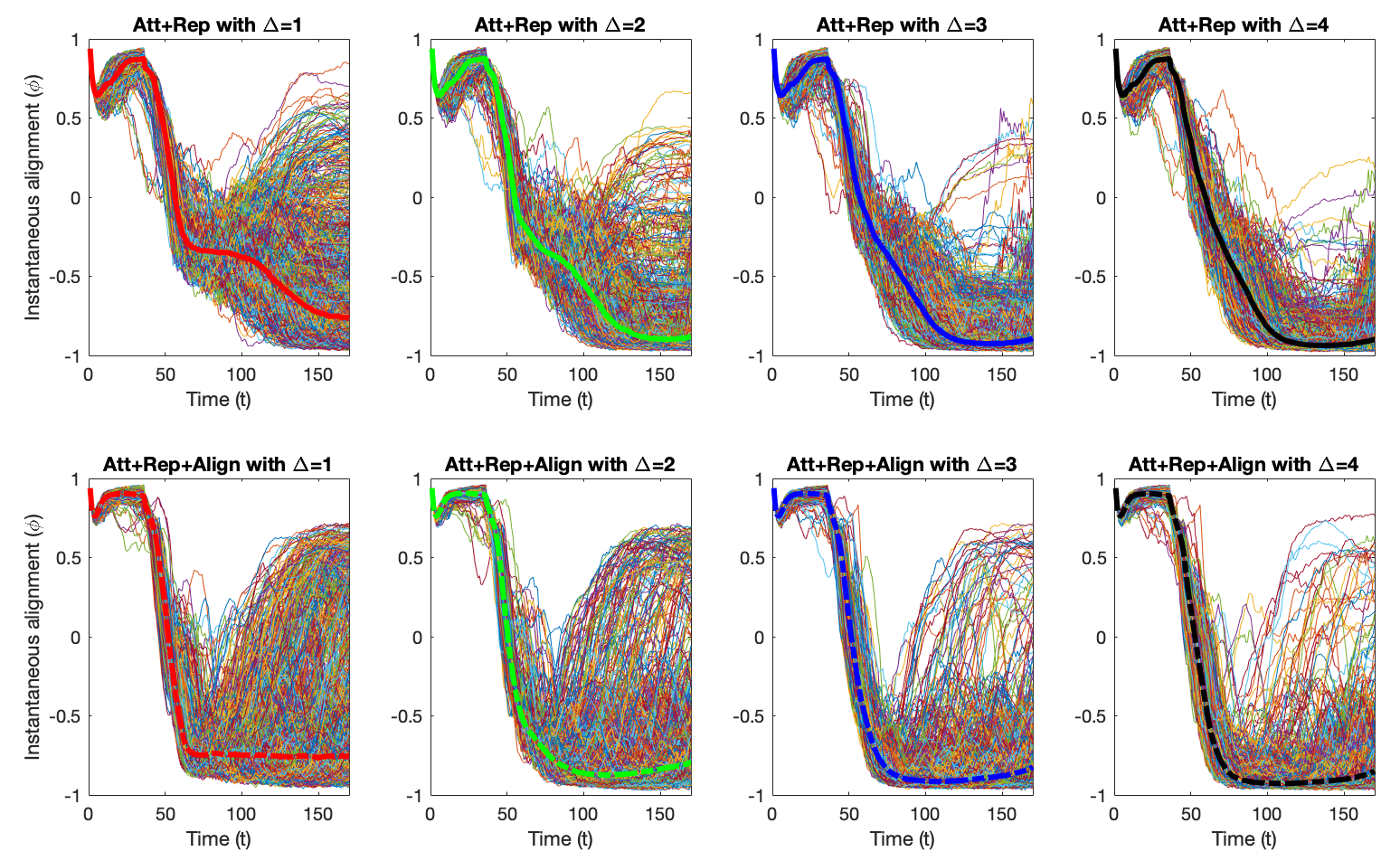}
\caption{Figure S2: Measured instantaneous alignment in each simulation. Each panel shows the measured instantaneous alignment over time in each of the 1450 simulations for that case, with the median superimposed as thicker line. \label{fig4}}
\end{figure} 
\newpage
Video S1: Example simulations for each of the eight cases.

\end{document}